\documentclass[final,3p,times,twocolumn]{elsarticle}

\biboptions{sort&compress}

\usepackage[T1]{fontenc}
\usepackage[utf8]{inputenc}
\usepackage{amsmath,amssymb}
\usepackage{booktabs}
\usepackage{siunitx}
\usepackage{xspace}
\usepackage[hidelinks]{hyperref}

\newcommand{\ares}{\textsc{ARES}\xspace}
\newcommand{\qe}{\textsc{Quantum ESPRESSO}\xspace}

\begin{document}

\begin{frontmatter}

\title{ARES: a plane-wave density-functional-theory code in Rust}


\author[1,2]{Zehan Liu}
\author[1,2]{Qian Wang}
\author[1,2]{Xiangyu Zheng}
\author[1,2]{Jiaxiang Li}
\author[1,2]{Chuntian Wu}
\author[1,2]{Yongshuo Chen}
\author[1,2]{Xuecheng Shao}
\author[1,2]{Qiang Xu}
\author[1,2]{Yu Xie\corref{cor1}}
\ead{xieyu@jlu.edu.cn}
\author[1,2]{Yanming Ma\corref{cor1}}
\cortext[cor1]{Corresponding authors.}
\affiliation[1]{organization={Key Laboratory of Material Simulation Methods
                              and Software of Ministry of Education,
                              Jilin University},
                city={Changchun},
                postcode={130012},
                country={China}}
\affiliation[2]{organization={Jilin University},
                city={Changchun},
                postcode={130012},
                country={China}}

\begin{abstract}
\noindent
\ares{} (\emph{Ab initio} atomic mateRial modEling Software at JLU) is a
plane-wave density-functional-theory (DFT) code written entirely in Rust as
a layered 22-crate workspace. It supports norm-conserving pseudopotentials
and the projector augmented-wave (PAW) method with the local-density
approximation (LDA) and Perdew--Burke--Ernzerhof (PBE) functionals,
$\Gamma$-only and explicit $k$-point sampling, smearing and mixing, forces,
stress, relaxation, band structures, densities of states (DOS) and
dispersion corrections; execution is serial or MPI-distributed over
$k$-point pools, band groups and plane-wave rows.

The code's distinguishing engineering choice is its validation apparatus.
Every numerical component is checked against \qe{} driven from the same
physical and numerical inputs, and the comparison is organized as numbered
fixtures kept in the repository whose reference data is regenerated by a
script instead of being captured once. On six materials spanning a
semiconductor, three metals and two two-dimensional systems, band energies
within one electronvolt of the Fermi level agree with the reference to a
median below \SI{0.4}{\milli\electronvolt}. The pseudopotentials the code reads are validated against all-electron
reference data as well: the median delta factor is
\SI{0.696}{\milli\electronvolt\per atom} over all 65 paired elements. Restrictions
are stated with the capabilities:  two functionals, no
spin-polarized or spin--orbit path, a stress implementation with an explicit
scope, and no accelerator execution.
\end{abstract}

\begin{keyword}
density functional theory \sep plane-wave pseudopotential \sep
projector augmented wave \sep Rust \sep software validation \sep
reproducibility
\end{keyword}

\end{frontmatter}


\section{Introduction}

\label{sec:intro}

Electronic-structure theory reaches its users almost entirely through
large, long-lived software packages written predominantly in Fortran:
\qe{}, ABINIT, SIESTA, CONQUEST, GPAW, ABACUS, DFTB+, SPARC, JDFTx,
exciting and BigDFT~\citep{Giannozzi_2020,Gonze_2009,Garc_a_2020,
Nakata_2020,Mortensen_2024,Zhou_2025,Hourahine_2025,Xu_2021,
Sundararaman_2017,Gulans_2014,Raya_Moreno_2026,Mohr_2014}. These codes have accumulated decades of
validated physics and are now expected to absorb hardware generations their
architectures were not designed for, a pressure visible in their own
reporting documents~\citep{Giannozzi_2020,Carnimeo_2023}. The
current generation of status reports shows how much of the development
budget now goes into that adaptation: Abinit 2025, the FHI-aims roadmap
and hybrid functionals in
SIESTA~\citep{Verstraete_2025,Abbott_2026,Pouillon_2026}. The performance frontier of established
plane-wave codes continues to move as well~\citep{Yao_2026}. Successive community roadmaps
quantify both the requirement and the cost: portability to heterogeneous
nodes, shrinking developer pools for legacy dialects, and the difficulty of
introducing new numerical strategies without perturbing validated
pathways~\citep{Gavini_2023,Blum_2024}.

A distinct line of work re-implements the codes instead of retrofitting
them, on the premise that the implementation language determines which
software architectures are practical. DFTK~\citep{Herbst2021} and
PWDFT.jl~\citep{Fathurrahman_2020} exploit Julia's dispatch and
composability, a route recently taken further by the plane-wave code
Hylanemos~\citep{Jie_2025}; PyPWDFT~\citep{Gao_2025,Gao_2026} pursues a
lightweight Python single-node plane-wave code; INQ~\citep{Andrade_2021,Simoni_2025}
and MRChem~\citep{Wind_2022} are modern C++ rewrites oriented towards
accelerators and multi-resolution representations. These codes go further
than porting: they restructure the software around language-level
guarantees.

Rust is an emerging candidate in this space. It offers systems-level
performance with compile-time exclusion of data races and no garbage
collector, which matters for iteratively parallelized SCF kernels, where
Fortran must rely on OpenMP conventions that the type system cannot
check~\citep{Jung_2017,Bitar_2024}. Rust is now established in
performance-critical scientific software beyond toy examples: MolAR, a
memory-safe trajectory-analysis library, benchmarks favourably against
established C/C++ tools~\citep{Yesylevskyy_2024}; a Rust-accelerated
diffraction kernel reports order-of-magnitude gains over its Python
predecessor in a materials-science workflow~\citep{Lebeda_2026}; and
dense linear algebra and HPC task
runtimes~\citep{Kazdadi_2026,Friese_2024} have peer-reviewed Rust
implementations. The published record in electronic structure itself is sparse. A search of
arXiv metadata, OpenAlex title and abstract phrases and Crossref titles,
combining Rust with plane-wave, pseudopotential, PAW or $k$-point
terminology, returns no software paper; the only recurring hit is
REST~\citep{Li_2025}, a molecular package with no plane-wave, periodic or
pseudopotential component. We therefore state, to our knowledge, that no
published paper whose title or abstract presents a Rust plane-wave periodic
DFT code exists.
The code described here, \ares{} (\emph{Ab initio} atomic mateRial
modEling Software at JLU), has been in production use for some time and
already appears in the literature, through an earlier method paper on
real-space Chebyshev-filtered subspace iteration~\citep{Xu_2019} and, for
example, in an automated machine-learning-potential
workflow~\citep{li2025selfoptimizingmachinelearningpotential}. What has been
missing is a description of the software itself, and this paper supplies
that description.

The code has grown into an ecosystem with companion modules written in
C/C++: a parallel atomistic simulation module providing


A third motivation is methodological. Electronic-structure codes are
routinely validated by physical benchmarks, but component-level numerical
records that can be re-run and regressed are rarely published, although
numerical cross-checking against an established code is increasingly used
as publishable evidence in its own right~\citep{Yang_2026}; changes to
precision or parallelism are usually justified a posteriori. \ares{} is
built the other way round: each numerical component is cross-validated
against a mature plane-wave implementation with identical inputs and
identical convergence parameters, the comparison is organized as numbered
fixtures kept in the repository, and precision-changing features are
admitted only through a pre-declared, layered acceptance protocol.

Two features distinguish \ares{} from the codes above. It is implemented
in Rust. It also exposes numerical precision to the user, where every code
in the list above fixes it at compile time or buries it inside a solver.

The remainder of this paper is organized as follows.
Section~\ref{sec:overview} gives a user-facing overview of what \ares{}
computes, how it is driven, and where its restrictions lie.
Section~\ref{sec:methods} describes the numerical methods and their
implementation. Section~\ref{sec:architecture} presents the software
architecture and engineering practices. Section~\ref{sec:validation}
reports the cross-validation corpus and the agreement it establishes. Section~\ref{sec:related} compares
\ares{} with related codes, and Section~\ref{sec:conclusions} concludes.



\section{Overview: what \ares{} is and what it computes}
\label{sec:overview}

\ares{} is a plane-wave pseudopotential code for periodic solids. It solves
the Kohn--Sham equations~\citep{Hohenberg_1964,Kohn_1965} self-consistently
for a periodic cell and, from the
converged state, produces forces, stress and the electronic-structure
quantities that follow from the density and the eigenvalues. It is written
in Rust and organized as a workspace of 22 crates, with the numerical core,
the parallel layer, the input and output layer and six native bridges kept
in separate crates.

The physical scope is the one this combination of basis and pseudopotential
covers: crystalline and molecular solids described by a periodic cell, with
norm-conserving or projector augmented-wave pseudopotentials read from UPF
files, and with the local-density or Perdew--Burke--Ernzerhof functional.
The code computes ground-state energies with their
component decomposition, charge densities, eigenvalues and occupations, and
from the converged state it computes forces, stress tensor elements,
fixed-cell and variable-cell optimized geometries, band structures along
high-symmetry paths, total and orbital-projected densities of states, and
dispersion-corrected energies. Section~\ref{sec:methods} describes how each
of these is implemented. Sampling is $\Gamma$-only or by an explicit set of
$k$ points, with automatic mesh generation and symmetry reduction to the
irreducible wedge.

Runs are serial by default and scale through MPI over three axes: pools that
own independent $k$ points, band groups that divide the column work of the
iterative solver, and groups that divide plane-wave rows and the slabs of a
distributed fast Fourier transform. A subset of kernels also uses Rayon
threads. Both are optional build features, and a binary compiled with MPI
still runs a serial calculation unless a process grid is configured.

\subsection{Input, workflow and state reuse}
\label{sec:overview:workflow}

An input has three parts: a control file, a structure file and one
pseudopotential file per species. The control file is organized in named
blocks covering the calculation to run, the cell and species, symmetry and
$k$-point sampling, the electronic loop, the ionic and cell degrees of
freedom, and output options. Relative paths resolve against the directory
that holds the control file, so an input directory stays relocatable.

Producers and consumers are separated across processes. A self-consistent
run, a fixed-cell relaxation and a variable-cell relaxation each write a
canonical state directory holding the density, the PAW data, the Fermi
energy and provenance information. The band-structure, density-of-states and
projected-density-of-states commands read that directory through the same
output-directory and prefix settings, and none of them overwrite the state
they read. The state directory is assembled in a temporary directory and
moved into place, so an interrupted run leaves no partial state behind.

\begin{table*}[t]
  \centering
  \caption{The command-line surface. The thirteen entries divide into four
           groups: the dispatcher, three informational commands that run no
           calculation, an offline export of a saved state, and the seven
           commands that solve or diagonalize. Every command except the
           dispatcher and the two dump tools takes the same control file,
           and the offline tools are the only ones that do not require the
           linear-algebra feature, so a build without it exposes a smaller command set
yinstead of failing at run time. Molecular
           dynamics is listed in the control-file grammar and is rejected
           when it is selected.}
  \label{tab:commands}
  \small
  \setlength{\tabcolsep}{5pt}
  \begin{tabular}{llll}
    \toprule
    Command & Consumes & Produces & Needs \texttt{lapack} \\
    \midrule
    \multicolumn{4}{l}{\emph{Dispatch and information}} \\
    \texttt{ares <control>}      & control file & whichever task it names & yes \\
    \texttt{ares --help}         & nothing & usage text & no \\
    \texttt{ares --backends}     & nothing & compiled numerical backends & no \\
    \midrule
    \multicolumn{4}{l}{\emph{Input and structure tools}} \\
    \texttt{dump-input}          & control file & canonical resolved input to stdout & no \\
    \texttt{dump-atoms}          & structure file & parsed cell and species to stdout & no \\
    \texttt{export}              & control file & POSCAR, VASP, XSF or ARES state in a new directory & no \\
    \midrule
    \multicolumn{4}{l}{\emph{Solvers and diagonalization}} \\
    \texttt{scf}                 & control file & canonical state directory & yes \\
    \texttt{nscf}                & control file & canonical state directory & yes \\
    \texttt{bands}               & converged state & eigenvalues and path coordinates & yes \\
    \texttt{dos}                 & converged state & density of states on an energy grid & yes \\
    \texttt{pdos}                & converged state & projected densities of states & yes \\
    \texttt{relax}               & control file & optimized geometry and final state & yes \\
    \texttt{vc-relax}            & control file & optimized cell, geometry and final state & yes \\
    \bottomrule
  \end{tabular}
\end{table*}

\subsection{Building and running}
\label{sec:overview:build}

The code builds with a stable Rust toolchain and needs a BLAS and LAPACK
provider plus \texttt{pkg-config}. Optional capabilities are compile-time
features: the LAPACK bridge, which every solving command requires; MPI and
Rayon as separate features; ScaLAPACK, \texttt{libxc} and
\texttt{simple-dftd3}, each behind its own flag; an alternative vendored FFT
backend; and a CUDA feature that supplies pinned-host and device memory
spaces without connecting any GPU to the numerical path.

Verification is a documented set of local commands: a workspace check, the
test suite, a linter run with warnings denied, and feature-specific test
invocations. Continuous integration is not configured in the repository.

\section{Theory and numerical methods}
\label{sec:methods}

This section states what the code implements and fixes the conventions that
later sections depend on. The approach is the standard plane-wave
pseudopotential one~\citep{Payne_1992}; the underlying theory is cited
where it is used. It is written out only where a citation would leave a
reader unable to reproduce a reported number.

\subsection{Plane-wave basis and grids}
\label{sec:methods:basis}

Wavefunctions are expanded in plane waves over the reciprocal cell,
\begin{equation}
\psi_{n\mathbf{k}}(\mathbf{r}) = \frac{1}{\sqrt{\Omega}}
\sum_{\mathbf{G}} c_{n\mathbf{k}}(\mathbf{G})\,
e^{\,i(\mathbf{k}+\mathbf{G})\cdot\mathbf{r}},
\label{eq:basis}
\end{equation}
with the sum restricted to plane waves whose kinetic energy is below the
wavefunction cutoff $E_{\mathrm{wfc}}$. Because the kinetic energy of a
plane wave is $\tfrac{1}{2}|\mathbf{k}+\mathbf{G}|^{2}$ in hartree, that
restriction fixes the basis size. The real-space grid is set by the density
cutoff instead, which defaults to four times the wavefunction cutoff but can
be given independently. The number of
plane waves grows with the cell volume and with the cutoff as
\begin{equation}
N_{\mathrm{PW}} \;\simeq\; \frac{\Omega}{6\pi^{2}}
\left(2E_{\mathrm{wfc}}\right)^{3/2},
\label{eq:pwcount}
\end{equation}
which is why the cost of a calculation rises steeply with the cutoff and
only linearly with the number of atoms at fixed density of $k$ points. The
expression counts the full sphere, while a calculation that samples only the
zone centre stores half of it and a shifted mesh stores a sphere displaced by
$\mathbf{k}$, so it is an estimate of the growth and not a
count of the stored coefficients. The dense transform box resolves the density. Its extent along each
reciprocal direction follows from the largest Miller index inside the
cutoff sphere. The projection of the sphere radius onto a lattice vector
bounds that index from above, and the bound is used as the search window and
not as the answer, because in a skewed cell the largest index
inside the sphere is strictly smaller than the projection,
\begin{equation}
\begin{split}
n^{\max}_{i} &\le \left\lfloor
\frac{\sqrt{2E_{\rho}}\,|\mathbf{a}_{i}|}{2\pi} \right\rfloor , \\
N_{i} &= \min\left\{\, n \ge 2n^{\max}_{i} + 1 \;:\;
n = 2^{p}3^{q}5^{r} \,\right\},
\end{split}
\label{eq:fftdims}
\end{equation}
with $\mathbf{a}_{i}$ the direct lattice vectors. Restricting the prime
factors to $2$, $3$ and $5$ keeps the transform on its fast paths at the
cost of a few percent more grid points than the minimum. The largest Miller
index is taken by enumeration instead of from the bound above, so the
dimensions are exact for a skewed cell.

Two further cutoffs define the set used for the density: a fine set bounded
by $E_{\rho}$, and a smooth set bounded by
\begin{equation}
E_{\mathrm{smooth}} = \min\!\left(E_{\rho},\,4E_{\mathrm{wfc}}\right),
\label{eq:smooth}
\end{equation}
which is active only when $E_{\rho} > 4E_{\mathrm{wfc}}$. Three compact
reciprocal-space sets therefore coexist, and the charge density and the
augmentation charges move between them by zero padding on the dense box.
When $E_{\rho}$ is not supplied it defaults to $4E_{\mathrm{wfc}}$, and the
smooth set coincides with the fine set. The transforms use a Rust FFT
library by default, with an alternative vendor backend selectable at build
time; both are driven through one plan interface.

The charge density is built from the occupied states,
\begin{equation}
\rho(\mathbf{r}) = \sum_{n\mathbf{k}} w_{\mathbf{k}}\, f_{n\mathbf{k}}\,
\left|\psi_{n\mathbf{k}}(\mathbf{r})\right|^{2},
\label{eq:density}
\end{equation}
where $w_{\mathbf{k}}$ is the $k$-point weight multiplied by the spin
degeneracy of a non-spin-polarized calculation, so that it sums to two over
the Brillouin zone, and $f_{n\mathbf{k}}$ are the occupations of
Section~\ref{sec:methods:scf}. The weight the code stores, symmetrizes and
writes out is the geometric one and sums to one; the factor of two is applied
where it is used, here and in Eq.~\eqref{eq:rhoij}, so a reader matching
these equations against a stored weight should expect that factor. The density is the quantity
the self-consistent cycle iterates on. It is accumulated in double precision
from the wavefunction coefficients, without a compensated summation.

Sampling is either $\Gamma$-only or an explicit list of $k$ points. Meshes
are generated by the Monkhorst--Pack construction~\citep{Monkhorst_1976}
and carry uniform weights. The code does not shift or prune a mesh on its
own. Symmetry operations are detected from the cell metric, and the mesh is
reduced to the irreducible wedge; the density and the augmented-wave
occupation matrices are symmetrized back over the full mesh before the
effective potential is rebuilt from them.

\subsection{Pseudopotentials and the augmented-wave transformation}
\label{sec:methods:pseudo}

Pseudopotentials are read from UPF files~\citep{Giannozzi_2020}. Norm-conserving potentials and
projector augmented-wave datasets are both supported, including the
nonlinear core correction, which enters the exchange--correlation energy and
potential, the forces and the stress. The ultrasoft type is recognized in
the file format and has no production path.

The augmented-wave method lifts the norm-conserving constraint on the
valence partial waves by reconstructing the all-electron wavefunction from a
smooth one; the core stays frozen, as it does in the norm-conserving path,
\begin{equation}
|\psi_{n\mathbf{k}}\rangle =
|\tilde{\psi}_{n\mathbf{k}}\rangle
+ \sum_{i} \bigl(
|\phi_{i}\rangle - |\tilde{\phi}_{i}\rangle
\bigr) \langle \tilde{p}_{i} | \tilde{\psi}_{n\mathbf{k}} \rangle ,
\label{eq:paw-transform}
\end{equation}
where $\phi_{i}$ and $\tilde{\phi}_{i}$ are the all-electron and
pseudo partial waves of the dataset and $\tilde{p}_{i}$ the projector
functions~\citep{Blochl_1994}. The transformation acts on the density
through a compensation charge that restores the correct multipole moments
of each augmentation sphere,
\begin{equation}
\hat{n}(\mathbf{r}) = \tilde{n}(\mathbf{r})
+ \sum_{ij} \rho_{ij}\, Q_{ij}(\mathbf{r}),
\label{eq:augcharge}
\end{equation}
with the occupation matrix formed from the projectors,
\begin{equation}
\rho_{ij} = \sum_{n\mathbf{k}} w_{\mathbf{k}}\, f_{n\mathbf{k}}\,
\langle \tilde{\psi}_{n\mathbf{k}} | \tilde{p}_{i} \rangle
\langle \tilde{p}_{j} | \tilde{\psi}_{n\mathbf{k}} \rangle .
\label{eq:rhoij}
\end{equation}
The expansion of $Q_{ij}$ in angular multipoles is carried on radial grids,
and the implementation checks the table's layout against
$n_{\mathrm{proj}}^{2}(2l_{\max}+1)$ while the dataset is parsed, so an
inconsistent table is rejected at that point. The table is stored, not contracted:
the kernels build $Q_{ij}$ from the radial partial waves and the tabulated
multipoles are not read again.

The nonlocal operator follows from the same data. Its strength is not a
constant of the dataset: one term follows the effective potential of the
current step and another follows the density through the projections, so the
coefficients are reassembled at every step,
\begin{equation}
D_{ij} = D^{(0)}_{ij}
+ \int V_{\mathrm{eff}}(\mathbf{r})\, Q_{ij}(\mathbf{r})\,
d^{3}r + \Delta_{ij} ,
\label{eq:dij}
\end{equation}
where $\Delta_{ij}$ is the one-centre exchange--correlation and Hartree
difference between the all-electron and pseudo partial waves inside the
augmentation spheres, taken with opposite signs for the two. That third
term is needed because the integral is evaluated with the pseudo effective
potential and not the all-electron one inside the spheres. The
generalized overlap operator that multiplies the pseudo wavefunctions is
\begin{equation}
S = 1 + \sum_{ij} |\tilde{p}_{i}\rangle
q_{ij} \langle \tilde{p}_{j}| ,
\label{eq:overlap}
\end{equation}
with $q_{ij}$ the integrals of the compensation charge. The overlap matters
beyond the eigenproblem: it is the operator the projected densities of
states are built with in Section~\ref{sec:methods:postproc}, so a
norm-conserving and an augmented-wave dataset go through one projection
rule. The one-centre energy is computed separately and reported as its own
term, which is what lets the decomposition of
Eq.~\eqref{eq:etot} be checked against a reference component by component.

\subsection{Exchange and correlation}
\label{sec:methods:xc}

Two functionals are implemented natively. The local-density approximation
combines Slater exchange~\citep{Slater_1951} with the PW92 correlation
functional~\citep{Perdew_1992}, and the generalized-gradient functional is
the Perdew--Burke--Ernzerhof form~\citep{Perdew_1996} with both parts
implemented in the code. Naming the functional does not pin it down, because
the revised variants of it share the functional form and differ only in two
constants. The set implemented here is the original one: $\kappa = 0.804$
for exchange, $\beta = 0.0667$ for correlation, and
$\mu = \beta\pi^{2}/3 = 0.2195$ fixed by the cancellation of the exchange
and correlation linear responses. Variants that revise those numbers are
not reachable, so a calculation cannot be run under a name that suggests
one of them while using another. Each routine returns the energy density and the
exchange--correlation potential in one pass,
\begin{equation}
V_{\mathrm{xc}}(\mathbf{r}) =
\frac{\delta E_{\mathrm{xc}}[\rho]}{\delta \rho(\mathbf{r})},
\label{eq:vxc}
\end{equation}
so that the energy and the operator cannot drift apart through separate
evaluations. The gradient-corrected pair also needs
$|\nabla\rho|$ and its functional derivative on the dense grid, which is
the step that dominates the cost of this part. Both functionals have a
distributed variant that assigns the dense grid to the ranks of a
plane-wave communicator. Bindings to \texttt{libxc}~\citep{Lehtola_2018}
expose the same two pairs through a configuration identifier. No other
functional is reachable, and an unrecognized name is rejected while the
input is read. The comparison between the native implementation and the
library is settled for the local-density pair and open for the gradient
pair, where the two differ in their treatment of the correlation
functional.

\subsection{The Hamiltonian and its energy decomposition}
\label{sec:methods:hamiltonian}

The Kohn--Sham operator that the solver applies is
\begin{equation}
\begin{split}
\hat{H} = {} & -\tfrac{1}{2}\nabla^{2}
+ V_{\mathrm{loc}}(\mathbf{r}) + V_{\mathrm{H}}(\mathbf{r})
+ V_{\mathrm{xc}}(\mathbf{r}) \\
& + \sum_{ij} |\beta_{i}\rangle D_{ij}
\langle \beta_{j}| ,
\end{split}
\label{eq:hamiltonian}
\end{equation}
where the local term is taken from the pseudopotential in reciprocal space
and interpolated onto the dense grid, the projectors $\beta_{i}$ and the
coefficients $D_{ij}$ carry the nonlocal and one-centre contributions of
Eq.~\eqref{eq:dij}, and the augmented-wave overlap of
Eq.~\eqref{eq:overlap} accompanies it. The Hartree solver and the
exchange--correlation assembly both have distributed variants. Ewald
summation~\citep{Ewald_1921} supplies the electrostatic energy of the
periodic ion array, decomposed into real-space, reciprocal-space, self and
background contributions, together with the corresponding forces and
stress.

The reported total energy is decomposed into terms a reader can check one at
a time,
\begin{equation}
E_{\mathrm{tot}} = E_{\mathrm{band}} + E_{\mathrm{H}} + E_{\mathrm{xc}}
+ E_{\mathrm{Ewald}} - E_{\mathrm{dc}} + E_{\mathrm{ent}}
+ E_{\mathrm{1c}} + E_{\mathrm{disp}},
\label{eq:etot}
\end{equation}
where $E_{\mathrm{band}} = \sum_{n\mathbf{k}}
w_{\mathbf{k}} f_{n\mathbf{k}} \epsilon_{n\mathbf{k}}$ is the sum of the
occupied eigenvalues, the double-counting term
\begin{equation}
E_{\mathrm{dc}} = \int \bigl(V_{\mathrm{H}}(\mathbf{r})
+ V_{\mathrm{xc}}(\mathbf{r})\bigr)\, \rho(\mathbf{r})\, d^{3}r
\label{eq:edc}
\end{equation}
removes the potential energy that the band energy has already counted;
$E_{\mathrm{ent}}$ is the smearing entropy, which converts the band energy
into the free energy of Mermin~\citep{Mermin_1965} and vanishes only for
fixed occupations; $E_{\mathrm{1c}}$ is the energy of the one-centre
contributions inside the augmentation spheres and is present only on the
augmented-wave path; and the dispersion term is present only when a
dispersion correction is requested. The augmented-wave path replaces the
integrand of Eq.~\eqref{eq:edc} by the difference between the effective and
local potentials, and adds the matching augmentation sum to the same term.
The terms that have an operator in Eq.~\eqref{eq:hamiltonian} can be checked
against it one at a time, so a disagreement with a reference calculation can
be traced to that operator instead of only to the total.

\subsection{The single-particle problem and the SCF cycle}
\label{sec:methods:scf}

The eigenproblem is solved by an iterative Davidson
scheme~\citep{Davidson_1975}. Writing the
subspace basis as $X$, the coefficient matrix as $Y$ and the Ritz values as
$\Lambda$, each step forms the residual of the Ritz vectors $XY$,
\begin{equation}
R = HXY - SXY\Lambda ,
\label{eq:residual}
\end{equation}
preconditions it, and extends the subspace with the preconditioned
directions, after which the projected problem
\begin{equation}
\left(X^{\dagger} H X\right) Y
= \left(X^{\dagger} S X\right) Y \Lambda ,
\label{eq:ritz}
\end{equation}
is solved for the updated coefficients. The initial subspace is
orthonormalized by a Cholesky-QR factorization~\citep{Stathopoulos_2002};
a restart rotates the Ritz columns into the new basis and does
not re-orthonormalize, and trial directions whose norm falls below a rank
tolerance of $10^{-12}$ are discarded before they enter the subspace.
Convergence is measured on the change of the Ritz values between iterations.
For an occupied band the threshold is the requested one; for an empty band it
is floored at $\max(5\varepsilon_{\mathrm{thr}}, 5\times10^{-6}~\mathrm{Ha})$,
so that the empty states stop at a level the iterative solver can still
resolve, and are not driven to the limit set for the occupied ones. The
distinction needs three qualifiers. A band counts as occupied when its
occupation divided by the $k$-point weight exceeds $0.01$, and the
classification is read from the previous iteration, so on the first iteration
of a self-consistent run every band takes the requested threshold. A
non-self-consistent solve, a band structure or a density of states supplies no
classification at all and behaves the same way. The floor applies only where
the eigensolver is given the classification, which is the path built with the
dense linear algebra feature.

New directions are preconditioned with the Teter--Payne--Allan diagonal
approximation~\citep{Teter_1989}. With
$x = (h_{ii} - \epsilon_i s_{ii}) \cdot 2$ formed from the diagonal elements
of the Hamiltonian and overlap in the reciprocal basis, the doubling being the
conversion to the rydberg units the form is defined in, the denominator is
\begin{equation}
d_i = \tfrac{1}{2}\left(1 + x + \sqrt{1 + (x-1)^2}\right),
\label{eq:tpa}
\end{equation}
This denominator is bounded below by one, so it cannot vanish; the
single-precision evaluation is nevertheless paired with a double-precision
path taken once $x$ approaches the single-precision range, which keeps a state
near the cutoff from overflowing it.

The occupations that enter Eq.~\eqref{eq:density} are either fixed or
smeared with a Gaussian~\citep{Fu_1983}, in which case
\begin{equation}
f(\epsilon) = \frac{1}{2}\left[
1 - \mathrm{erf}\!\left(
\frac{\epsilon - \epsilon_{\mathrm{F}}}{\sigma}\right)\right],
\label{eq:smearing}
\end{equation}
where $\sigma$ is the broadening width taken unchanged from the input field
\texttt{degauss}, in hartree, with the Fermi level $\epsilon_{\mathrm{F}}$
found by requiring the
occupations to sum to the electron count. Other smearing schemes appear in the input grammar and are not wired to the
self-consistent solver. The density of states accepts one of them, the
Methfessel--Paxton form; the remaining two are rejected by both paths, as
Section~\ref{sec:methods:postproc} notes.

The charge density is mixed between iterations by one of four schemes:
linear mixing, linear mixing with a Thomas--Fermi preconditioner, and a
modified Broyden~\citep{Broyden_1965} mixer with and without that
preconditioner. The input keyword that selects the first two,
\texttt{mixing\_mode = "plain"} and \texttt{"tf"}, reaches the Broyden
mixer in both cases, with and without the preconditioner; the two linear
schemes are reachable only through the library entry point, which is worth
knowing before reading a run's convergence behaviour as that of a linear
mixer. Each scheme builds a new density from the current one and
the residual $\Delta\rho = \rho_{\mathrm{out}} - \rho_{\mathrm{in}}$,
\begin{equation}
\rho^{(m+1)} = \rho^{(m)} + \alpha\,
\mathcal{P}\!\left[\Delta\rho^{(m)}\right],
\label{eq:mix}
\end{equation}
where $\alpha$ is the mixing factor, every quantity is taken at the same
reciprocal vector $\mathbf{G}$, the preconditioner annihilates the
$\mathbf{G} = 0$ component so the mean density is left alone, the Broyden
coefficients act on the smooth set while the high-$\mathbf{G}$ tail keeps a
plain linear update at the same $\alpha$, and $\mathcal{P}$ is the identity for
linear mixing, the Thomas--Fermi preconditioner
\begin{equation}
\begin{split}
\mathcal{P}_{\mathrm{TF}}\!\left[\Delta\rho\right]
&= \frac{|\mathbf{G}|^{2}}{|\mathbf{G}|^{2} + k_{\mathrm{TF}}^{2}}\,
\Delta\rho, \\
k_{\mathrm{TF}}^{2} &= \frac{(12/\pi)^{2/3}}{r_{s}},
\qquad
r_{s} = \left(\frac{3\Omega}{4\pi N_{e}}\right)^{1/3},
\end{split}
\label{eq:tf}
\end{equation}
when that option is selected, and the Broyden update assembled from the
stored history when the Broyden mixer is used. The history length is fixed
at eight.

One further correction belongs to the iteration and not to
Eq.~\eqref{eq:etot}. Because the electronic step evaluates the band energy
and the potential terms at different density stages, the trace prints a
variational correction alongside the total: the integral of the difference
between the input and the output density against that step's Hartree and
exchange--correlation potential, with the matching augmented-wave term. It
vanishes once the density stops changing, and it is what makes the
per-iteration energy agree with the mixed density actually carried forward.
The consequence is worth stating: the trace's total and the final reported
energy coincide only in the last iteration, and a reader comparing a run
against a reference may be holding one where the other is meant.

The convergence threshold is given in hartree in the input and compared
internally against a criterion expressed in rydberg,
\begin{equation}
4\pi\Omega \sum_{\mathbf{G} \neq 0}
\frac{\left|\Delta\rho_{\mathbf{G}}\right|^{2}}
{|\mathbf{G}|^{2}} < 2\,\varepsilon_{\mathrm{conv}},
\label{eq:conv}
\end{equation}
so a threshold of $10^{-6}$ in the input corresponds to $2\times10^{-6}$ in
the internal units. What is compared is a density residual, the Hartree
energy of the difference between the input and the output density of the
step, which is read as an estimate of the error in the energy and
is not one. The reciprocal vectors entering the sum are those of the
smooth set of Eq.~\eqref{eq:smooth} on the paths the input can select, so
the reported residual is a smooth-set quantity and not a fine-grid one. A run stops when the criterion is met or when the step
limit is reached.

\subsection{Forces, stress and structure optimization}
\label{sec:methods:mechanics}

Forces are the derivative of the total energy with respect to the ionic
positions,
\begin{equation}
\mathbf{F}_{I} = -\frac{\partial E}{\partial \mathbf{R}_{I}},
\label{eq:force}
\end{equation}
and they are assembled term by term from the local, nonlocal, Ewald,
augmentation and core-correction contributions. A numerical derivative of
the total is never taken. They are rebuilt from the converged state instead of
from solver scratch, so the geometry criterion and the reported forces come
from the same wavefunctions. For the augmented-wave case the nonlocal contribution uses screened coefficients, because part of the
one-centre response is carried by the augmentation charge; the
two pieces are summed into a single reported nonlocal term. A correction
that accounts for the residual self-consistency error of the forces is
applied whenever forces are requested, with no input key that turns it off,
which matters when a calculation is deliberately stopped at a loose
threshold to save time. The
correction is added to the assembled terms, the mean force is then subtracted
so that the reported set carries no net drift, and the result is symmetrized
against the cell's point group.

The stress tensor is evaluated as the strain derivative of the energy at
fixed ionic coordinates,
\begin{equation}
\sigma_{\alpha\beta} = -\frac{1}{\Omega}
\left.\frac{\partial E}
{\partial \epsilon_{\alpha\beta}}\right|_{\epsilon=0},
\label{eq:stress}
\end{equation}
with the kinetic, Hartree, exchange--correlation, local, nonlocal and Ewald
contributions reported separately and then summed, the nonlinear core
correction folded into the exchange--correlation part and the augmentation
stress into the nonlocal one, the former on the augmented-wave path only,
since a norm-conserving calculation that carries a core correction is
refused rather than silently mis-stressed, and the dispersion contribution added when one
is requested. Its scope is narrower than that of the total energy, and
Section~\ref{sec:overview:limits} lists the combinations for which it is
available.

Geometry optimization runs a BFGS~\citep{Fletcher_1970} update on the ionic
coordinates at fixed cell. Variable-cell relaxation extends the same optimizer to the cell
degrees of freedom, with an optional external pressure and with constraints
on selected ionic coordinates.

\subsection{Post-processing}
\label{sec:methods:postproc}

Band structures are computed along a path through high-symmetry points. The
segment endpoints are kept exact, and the path coordinates are stored
alongside the eigenvalues so the path can be plotted without further
processing. Path weights never enter the occupations or the density: the
bands command reuses the converged potential and leaves the self-consistent
state untouched.

Densities of states are computed on a uniform energy grid from the
$k$-point sampling the input supplies, which is usually denser than the one
the self-consistent step used, taken on the same fixed potential,
\begin{equation}
g(E) = \sum_{n\mathbf{k}} w_{\mathbf{k}}\,
\delta\!\left(E - \epsilon_{n\mathbf{k}}\right),
\label{eq:dos}
\end{equation}
with the delta function
broadened by a Gaussian whose width is an input parameter, and the Fermi
level recomputed from that spectrum. The broadening for this command accepts the Gaussian and the
Methfessel--Paxton~\citep{Methfessel_1989} form, a wider set than the
self-consistent solver accepts for its occupations, because a density of
states is a post-processing quantity and does not feed back into the density.
The kernel is verified to tenth order, but the input grammar carries no order
field and the command is wired to the first-order form, so a reader cannot
reach a higher order from a control file. The other schemes named in the
grammar, Marzari--Vanderbilt and Fermi--Dirac, are rejected by this command
as well.

Projected densities of states use L\"owdin
projections~\citep{Lowdin_1950}. For each atomic orbital $\phi_{\mu}$ the
overlap matrix is formed with the generalized overlap operator $S$ of
Eq.~\eqref{eq:overlap},
\begin{equation}
O_{\mu\nu} = \langle \phi_{\mu} | S | \phi_{\nu} \rangle,
\qquad
\tilde{\phi}_{\mu} = \sum_{\nu}
\left(O^{-1/2}\right)_{\mu\nu} S \phi_{\nu},
\label{eq:lowdin}
\end{equation}
and the projected weight of band $n$ at $k$ is the squared projection onto
$\tilde{\phi}_{\mu}$,
\begin{equation}
p_{\mu n\mathbf{k}} =
\bigl| \langle \tilde{\phi}_{\mu} | \psi_{n\mathbf{k}} \rangle
\bigr|^{2}.
\label{eq:pdos}
\end{equation}
One projection rule therefore covers norm-conserving and augmented-wave
datasets alike. Projection inside the augmentation spheres is not
implemented, and the reported projections are the ones defined by
Eqs.~\eqref{eq:lowdin} and~\eqref{eq:pdos} alone.

\subsection{Dispersion corrections}
\label{sec:methods:dispersion}

The D2 correction~\citep{Grimme_2006} is implemented natively for periodic
cells. Pairs are generated by enumerating the lattice translations that fall
inside a spherical cutoff, and the energy is
\begin{equation}
\begin{aligned}
E_{\mathrm{D2}} &= -s_6 \sum_{i<j}
f_{\mathrm{damp}}\!\left(R_{ij}\right)
\frac{C_{6}^{ij}}{R_{ij}^{\,6}},\\
f_{\mathrm{damp}}(R) &=
\frac{1}{1 + e^{-20\left(R/R_0^{ij} - 1\right)}},
\qquad R_0^{ij} = R_0^{i} + R_0^{j},
\end{aligned}
\label{eq:d2}
\end{equation}
with the damping parameter fixed at the value used by the reference
implementation, $s_6$ taken from the input, and $C_6$ and the van der Waals
radii supplied per species, in $\mathrm{Ha}\,\mathrm{bohr}^{6}$ and bohr
respectively. The pair coefficient is the geometric mean of the two
per-species values, $C_{6}^{ij} = \sqrt{C_{6}^{i} C_{6}^{j}}$. The pair sum
runs over ordered pairs of atoms and periodic images, so an atom is paired
with its own images; the half that the loop carries on each ordered pair is
what makes the total equal the sum over $i<j$ written above, and only the
self term at zero displacement is skipped. The same pair list yields the Cartesian forces
and the stress. The D3 corrections, in the zero-damping and three-body
forms~\citep{Grimme_2010} and the Becke--Johnson
form~\citep{Becke_Johnson_2005}, come from a shared library loaded at run
time. Dispersion is not part of the self-consistent effective potential: it
is evaluated once per geometry and added to the energy, the forces and the
stress afterwards.

\subsection{Units and numerical conventions}
\label{sec:methods:units}

All internal state is held in hartree and bohr, and conversions happen at
the input and output boundary and nowhere else. The hartree--rydberg factor
is exactly two, which is what makes the cutoffs and the convergence
threshold comparable with those of other plane-wave codes after that factor.
The pseudopotential reader applies the same factor to the quantities stored
in rydberg in the UPF file, including the local potential, the coefficients
of the nonlocal projectors and the reference atomic charge. Energy and
length conversions use the CODATA values~\citep{Tiesinga_2021}, with one deliberate exception: the
constant converting an energy density to kilobar is retained at its legacy
value, about $3.7\times10^{-7}$ away in relative terms from the current
CODATA value, so pressures reported by long-standing variable-cell input
files keep their meaning.



\section{Software architecture and engineering}
\label{sec:architecture}

\subsection{Design constraints and language choice}
\label{sec:arch:rationale}

Four constraints shaped the implementation. The default numerical path has
to be reproducible, so that a change elsewhere in the code can be shown not
to have moved a number. The physics surface keeps growing, which puts a
premium on adding a feature without disturbing the validated paths around
it. Parallelism is optional and arrives in several forms, so the parallel
layer cannot be wired through the physics. Native libraries are needed for
linear algebra and for optional extras, and none of them should reach into
the numerical core.

The language arguments the project relies on are concrete. Ownership together with the \texttt{Send} and \texttt{Sync}
constraints excludes data races at compile time, which matters for kernels
that would otherwise depend on convention to stay correct under threads. The
reduction used in the threaded path is deterministic with respect to thread
count as a tolerance gate; it is not bitwise deterministic, and the code does
not claim otherwise. The input pipeline is enforced by typestate: the engine
accepts only a fully resolved configuration type, so passing an unresolved
one is a compile error instead of a runtime check. Deserialization is kept
pure, so configuration parsing performs no input or output and no cross-file
resolution; anything that depends on other files happens in the resolution


\subsection{Workspace layout and dependency rules}
\label{sec:arch:workspace}

The code is a Cargo workspace of 22 members: 16 crates under
\texttt{crates/} and 6 native bridges under \texttt{crates/ffi/}. The
numerical crates are layered, with the reciprocal-space and grid
construction, the pseudopotential reader, the fast Fourier transform, the
linear algebra, the Hamiltonian assembly and the self-consistent driver in
separate crates, and with the input and output layer, the command-line front
end and the profiling crate alongside them. Dependencies point downward,
and the architecture document names the directions that are forbidden: the
orchestrating crate may not depend on upper layers; the pseudopotential
reader may not depend on the Hamiltonian crate; the Fourier, linear-algebra,
parallel and memory crates may not depend on the self-consistent driver; and
the physics crates may not depend back on the driver or the front end. The native bridges sit outside that graph and are reached only through the
crates that own them; five of the six are reached from a numerical crate and
the sixth demonstrates the C++ interop mechanism without carrying physics.
The depth of a crate is the longest chain of dependencies below it, so nothing
is reachable from a crate that sits above it, and the bridges hang off the
outside where none of the forgotten directions can reach them.



\subsection{Parallel and memory infrastructure}
\label{sec:arch:parallel}

A distributed run places every rank on a three-axis grid. A rank carries the
coordinates $(p, b, g)$, and the total rank count equals the product of the
three extents: ranks along the first axis own independent $k$ points, ranks
along the second divide the column work of the iterative solver, and ranks
along the third divide plane-wave rows and the slabs of the distributed
transform. The extents come from the input or the command line, and the
relations among them and the rank count are checked when the grid is built.

Both parallel layers are optional. Ranks are placed with MPI, which is enabled by a build
feature and is not required, and threads come from a work-stealing runtime
behind a second feature. Neither is enabled by
default, and a binary compiled with MPI support runs a serial calculation
unless a process grid is requested. The ordinary self-consistent entry point
stays serial, which keeps the default path identical to the one used for the
regression comparisons described below.

Memory is handled through typed memory spaces. A sealed trait describes what a memory space must provide, aligned host arenas
are the default, and pinned-host and device spaces are compiled in only when
the corresponding optional feature is enabled. A device memory space of this
kind is storage and nothing more: it does not select an execution backend
and does not connect device memory to the solver. On top of the spaces sits
an arena with mark-and-release semantics, and buffers are accounted for
under one of four lifecycles: state that outlives a run, buffers reused
within a run, short-lived scratch, and memory that exists only to hold
reported or validation data. The distinction is recorded through the
profiling layer, so growth in the persistent category appears in the profile instead of having to be inferred.

\subsection{Foreign-function boundaries}
\label{sec:arch:ffi}

Native code is confined to the six bridge crates. Raw declarations and
native linkage stay inside them, and the rules the project enforces are that
high-level crates do not depend on raw interfaces, on C++ types, or on
handles that carry ownership, and that a Rust panic or a C++ exception never
crosses the boundary. Two of the bridges use a C++ interop layer, one loads
a shared library at run time, and the remainder wrap BLAS, LAPACK, ScaLAPACK and a vendor FFT backend. Only the
linear-algebra bridge is pulled in by the rest of the graph, and even that
dependency is gated: the command-line crate enables the solver crates only
when the linear-algebra feature is selected, which is why a build that omits
it has no commands that solve or diagonalize. Outside the bridges, unsafe
code is restricted to the memory-space layer of the backend crate, where it
is private and carries a documented invariant.

\subsection{Code size, tests and bit-identical regression}
\label{sec:arch:testing}

The workspace holds about 254\,000 lines of Rust in 603 files, of which
roughly 206\,000 lines are library and binary sources and the remainder are
tests. The largest crates are the self-consistent driver and the Hamiltonian
assembly, each near 60\,000 lines, followed by the one-step driver near
53\,000. Two crates are deliberately almost empty, holding a single numeric helper each, so that the layering has a designated place for such code. About
1\,800 test functions are distributed over the workspace, and the
reference corpus described in Section~\ref{sec:validation} lives under a
single test directory holding 72 fixture directories, 67 of them numbered,
with 764 stored reference files, all of them with a `.dat` extension.

The strongest of the regression mechanisms compares the default path
bit for bit. A recorder compiled only into tests writes the full 64-bit
representation of selected quantities to a file, driven by an environment
variable and requiring a named test thread, so the comparison is on raw
words and carries no tolerance. The acceptance runner that drives it pins
the thread-count environment of every numerical library, which is what makes
a repeated run comparable. A change that alters the order of operations of the default
path therefore fails the comparison outright, with no tolerance to absorb
it.

Verification is a documented set of local commands: a workspace check, the
test suite, a linter run with warnings denied, and feature-specific test
invocations for the optional capabilities. Continuous integration is not
configured in the repository, so these commands are run by hand, and the reproducibility claims in this paper rest on recorded outputs of those runs, since no automated pipeline exists.

\section{Validation and results}
\label{sec:validation}

This section reports how far the implementation can be trusted. The question
it answers is whether the code agrees with an established plane-wave
implementation when both are driven from the same physical and numerical
inputs, and whether the pseudopotentials it reads reproduce all-electron
reference data.

\subsection{The reference corpus}
\label{sec:val:corpus}

Reference values are produced by driving the reference code, Quantum
ESPRESSO 7.4, from the same input decks that the repository keeps, and
storing the results as expected data. The stored values are never read by
the code during a normal run. They live in 72 fixture directories, of which
67 carry a number, and comprise 764 reference data files. The numbering
runs from 018 to 082 with gaps and two duplicated numbers; the gaps are
historical and the numbering is not a coverage claim.

The fixtures are organized by component: transform and initialization, a
one-step non-self-consistent acceptance with and without the nonlocal terms,
augmented-wave initialization together with the Qij, rhoij and Dij updates,
self-consistent runs for each mixing mode, Davidson traces, converged
self-consistent runs for norm-conserving and augmented-wave datasets,
symmetry and irreducible-wedge reduction, forces including the nonlinear
core correction and the augmented-wave contributions, stress including the
local-density and augmented-wave cases, fixed-cell and variable-cell
relaxation, band structures, densities of states and projected densities of
states for both dataset types, dispersion corrections, the force correction
used for loosely converged self-consistent runs, and a multi-species
augmented-wave case.


\subsection{Component-level agreement}
\label{sec:val:components}

Agreement is reported component by component, which is the purpose of the
corpus: a single aggregate number would hide which part of the
implementation is responsible for a deviation.

The decomposed energy terms, the charge density and the augmented-wave
becsum agree with the reference to the level at which the two
implementations can be compared at all, meaning the arithmetic sets the difference and the algorithms do not. Converged augmented-wave runs
clear a \SI{0.1}{\milli\electronvolt\per atom} gate, and the broader
screening sweep over many systems uses \SI{1e-5}{\hartree\per atom}, about
\SI{0.272}{\milli\electronvolt\per atom}. Both numbers appear in the
repository and they apply to different sweeps.

Forces for norm-conserving datasets agree with the reference to
\SI{1.283e-6}{\hartree\per\bohr}, and against a finite-difference
derivative of the self-consistent energy to \SI{1.426e-6}{\hartree\per\bohr},
against an architecture goal of \SI{1e-4}{\hartree\per\bohr}. Reduction to
the irreducible wedge changes the forces by \SI{1.084e-5}{\hartree\per\bohr}
relative to an explicit full-mesh calculation, which is the size of the symmetrization effect, an artefact of the reduction itself. The symmetry of the reported
forces is checked against a \SI{5e-12}{\hartree\per\bohr} gate and is
observed at \SI{1.3e-14}{\hartree\per\bohr}. The nonlinear core correction
carries its own gate of \SI{2e-7}{}.

Stress for norm-conserving datasets matches the reference with total tensor
elements below \SI{2.5e-9}{\hartree\per\bohr\cubed} and raw components below
\SI{0.0055}{\kilo\bar}, and an independent strain finite-difference check
gives \SI{7.10e-10}{} and \SI{1.23e-9}{} for the two cases tested.

Dispersion corrections clear physical gates of \SI{1e-8}{}, \SI{1e-7}{} and
\SI{1e-8}{} on the energy, the forces and the stress, with observed
agreement below \SI{1e-12}{}. Band eigenvalues are compared against a formal
gate of \SI{5e-7}{\hartree}, and the observed differences are far smaller:
\SI{7.818e-10}{\hartree} and \SI{2.365e-8}{\hartree} on the two cases
reported, with a maximum band difference of \SI{8.2e-5}{\milli\electronvolt}
and a root-mean-square difference of \SI{8e-6}{\milli\electronvolt} for
the norm-conserving silicon case, and \SI{6.44e-4}{\milli\electronvolt} and
\SI{2.21e-4}{\milli\electronvolt} for the augmented-wave case.
Variable-cell relaxation clears the \SI{0.1}{\milli\electronvolt} gate with
\SI{1.78e-7}{\hartree} observed.

Densities of states are compared on the fixed energy grid the two codes share,
with the curve normalized because the reference stores it to limited precision.
Over the two cases, the Fermi level agrees to \SI{7.892e-6}{\hartree} and
\SI{6.071e-5}{\hartree} against a gate of \SI{1e-4}{\hartree}, the energy
grid to \SI{2.220e-16}{\hartree} against \SI{1e-12}{\hartree}, the
normalized density of states to \SI{4.681e-4}{} and \SI{4.633e-4}{} against
\SI{1e-3}{}, and the integrated density of states to \SI{4.982e-3}{} and
\SI{4.986e-3}{} states against \SI{6e-3}{}. The residual difference comes
from the spectrum and not from the accumulation grid or the broadening
kernel, since the grid agrees to machine precision while the density does not.

Projected densities of states need a different comparison, because degenerate
eigenstates are defined only up to a unitary rotation and a state-by-state
check would compare quantities that are not observable. The comparison
therefore sums channels within each degenerate cluster and checks the
per-state channel sums. The largest per-state channel-sum difference is
\SI{1.206e-4}{} for the norm-conserving case and \SI{6.792e-5}{} for the
augmented-wave case, and the largest per-channel difference inside a
degenerate cluster is \SI{4.152e-5}{} and \SI{2.145e-5}{}. The spilling is
a single number per calculation and therefore free of that ambiguity: it
agrees to the precision the reference reports, \num{0.00876387} against
\num{0.0088} and \num{0.00889913} against \num{0.0089}. Distributing the
calculation over ranks moves the normalized density of states by
\SI{3.697e-12}{} against a gate of \SI{1e-10}{}.


\subsection{Band structures, densities of states and pseudopotential accuracy}
\label{sec:val:bands}

The component checks above run on the fixture corpus, which is built around a
small number of systems. To test the electronic structure over a wider range of
physics, six materials were carried through a relaxation, self-consistent,
band-structure and density-of-states sequence: an indirect-gap semiconductor,
an alkali metal, a simple metal, a $d$-electron metal, a two-dimensional
semimetal and a two-dimensional semiconductor.

Agreement with another plane-wave code bounds one thing: that two
implementations of the same equations converge on the same answer. It says
nothing about how far that answer sits from the all-electron one, which is a
property of the pseudopotential. The delta factor measures that second
distance directly, and the two numbers together are what this section
reports.

\begin{figure}[!htbp]
  \centering
  \includegraphics[width=\columnwidth]{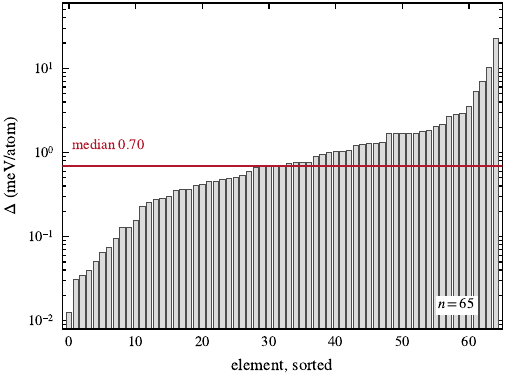}
    \caption{Delta factor per element for the JTH-PBE-v2.0 pseudopotential
           set, over the $65$ elements that produce a value on a converged
           run. The values are sorted on a logarithmic axis; the horizontal
           line marks the median and the worst case is labelled. The
           comparison is against WIEN2k, so no element is set aside for the
           $k$-point sampling another plane-wave code would choose.}
  \label{fig:validation}
\end{figure}

The delta factor~\citep{Lejaeghere_2016} compares the total energy of a
fixed set of crystals against WIEN2k~\citep{Blaha_2020} reference values.
It is therefore sensitive to the pseudopotential and to the functional,
while the geometry drops out. The calculation uses the JTH-PBE-v2.0
set~\citep{Jollet_2014} at the cutoffs listed in that set's own report. Of
65 paired elements all produce a value on a converged run, and
Figure~\ref{fig:validation} shows those sorted. The median delta factor is
\SI{0.696}{\milli\electronvolt\per atom}, the mean is
\SI{1.541}{\milli\electronvolt\per atom}, and the worst case is
\SI{22.97}{\milli\electronvolt\per atom}. The comparison is against the
WIEN2k numbers alone, so the $k$-point sampling another plane-wave code
would choose does not enter it and no element is set aside for a
disagreement on that axis. A reader comparing these values with a
cross-code table should know that 28 of the 65 pairs use different
$k$-point counts on the two plane-wave sides, up to a factor of five in the
most extreme case, and that those 28 are not the worse half: their median
is \SI{0.683}{\milli\electronvolt\per atom} against
\SI{0.762}{\milli\electronvolt\per atom} for the 37 that match.

The distribution matters more here than the median. Removing the single
worst case reduces the mean from \SI{1.541}{\milli\electronvolt\per atom}
to \SI{1.206}{\milli\electronvolt\per atom}, so the mean is not a useful
summary and the median is quoted as the headline. A small number of
transition-metal elements carry a delta factor an order of magnitude above
the rest, which is a property of those particular pseudopotentials and not of the code, and the repository records the same elements as known
difficulties of the set. We report the values as measured and do not
attribute them further without a citable source.

The pseudopotential set behind the delta numbers is the JTH-PBE-v2.0 table in
UPF format, which supplies 86 elements. Of those, 65 enter the paired comparison. The
library reader accepts norm-conserving and augmented-wave UPF datasets; it does
not read the alternative XML serialisation of the same table, which is a capability restriction and not a limit on the validation, and it is listed
with the other restrictions in Section~\ref{sec:overview:limits}.

\begin{table}[t]
  \centering
  \caption{Delta factor of the JTH-PBE-v2.0 set against the all-electron
           WIEN2k reference. All $65$ paired elements enter: the comparison
           is against WIEN2k alone, so an element is not set aside for the
           $k$-point sampling another plane-wave code would choose. The mean
           is quoted with and without the single worst element, because one
           value sits an order of magnitude above the rest and moves it.}
  \label{tab:pseudo}
  \footnotesize
  \setlength{\tabcolsep}{4pt}
  \begin{tabular}{lr}
    \toprule
    Quantity & $\Delta$ (meV/atom) \\
    \midrule
    Elements paired and used & $65$ \\
    Median                   & $0.696$ \\
    Mean                     & $1.541$ \\
    Mean without the worst   & $1.206$ \\
    Worst (Pd)          & $22.97$ \\
    \bottomrule
  \end{tabular}
\end{table}

Both codes are run on one geometry per material, the structure the reference
code relaxed to. Letting each code relax the cell and then computing bands at
its own result folds the difference between the two relaxations into the band
difference, and for the alkali metal those volumes differ by
\SI{1.79}{\percent}. On one geometry the band comparison is a statement about
the band implementation, which is what this section is for.

Each calculation is aligned to its own reference energy, the valence-band
maximum for a gapped system and the Fermi level of the density-of-states step
for a metal. Those two alignments agree to within
\SI{1.5}{\milli\electronvolt} for every material, so the choice is not carrying
the result. Aligning a gapped system on its Fermi level instead is a mistake and
not a refinement: that level follows from the broadening and
from the details of the solve, and the two codes place it
\SI{1.07}{\electronvolt} apart inside the same \SI{1.77}{\electronvolt} gap of
the two-dimensional semiconductor. Using it as the alignment would displace
every band by that amount.

\begin{figure*}[!t]
  \centering
  \includegraphics[width=0.98\textwidth]{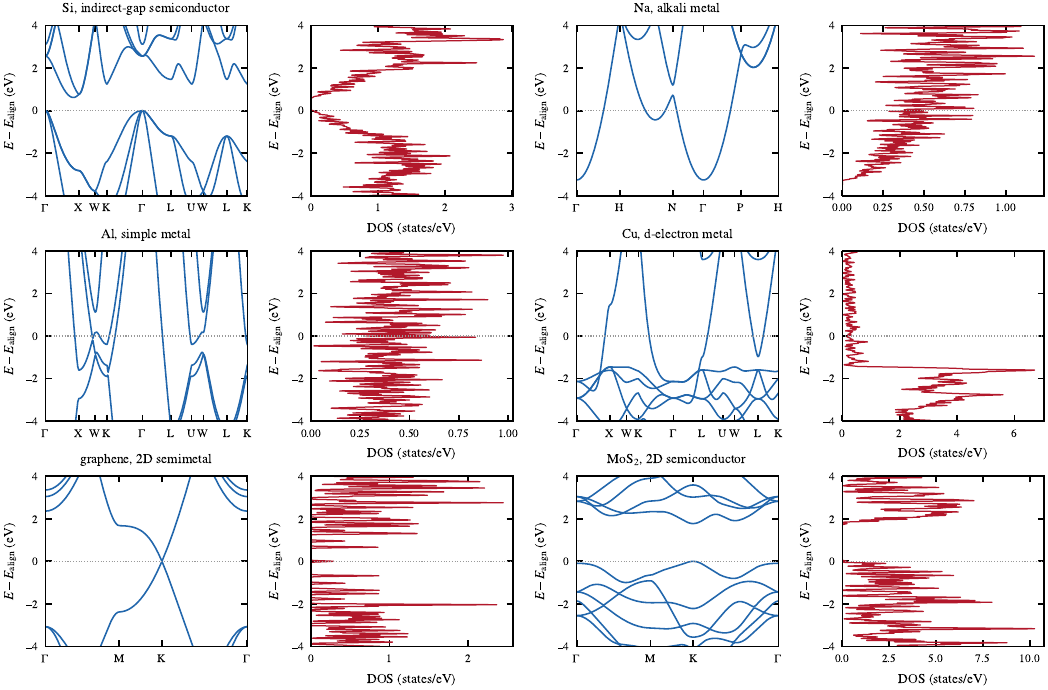}
  \caption{Band structure and density of states for the six materials, each
           panel pair computed in one pass at the geometry named in the text.
           Energies are referred to $E_{\rm align}$, the valence-band maximum
           for the gapped systems and the density-of-states Fermi level for the
           metals. The densities of states are drawn over the same
           \SI{4}{\electronvolt} window as the band structures, so a pair can be read as
           one picture; a state outside that window is
           not shown. The window is what keeps the panels comparable: the
           alkali metal carries a semicore peak two orders of magnitude larger
           than its valence density of states, and an axis scaled to include it
           would leave the region that decides the physics flat against the
           frame.}
  \label{fig:banddos}
\end{figure*}

The comparison covers the bands that decide the gap: the three highest occupied
bands and the three lowest empty ones. That restriction is a band index range
and not an energy range, and the reason is the solver and not the physics.
The benchmark this data comes from re-ran its reproducibility check against
the build reported here: two runs of an identical input, a Davidson subspace
enlarged from four to eight times the band count, and a hundredfold tighter
fixed-potential threshold all reproduce every band of every material in the
window to the last digit. The window is kept because the convergence guarantee
is stated for the states it admits; the earlier build did move the highest
empty bands by hundreds of milli\-electronvolts, so an energy window wide
enough to be interesting reaches into states whose convergence is not
promised.
Figure~\ref{fig:banddos} shows the band structures and densities of states
for all six materials as this code computes them, and
Table~\ref{tab:bands} differences the same bands against the reference
state by state. Over the band window the median difference is at most
\SI{0.2}{\milli\electronvolt} in any material and the root mean square at
most \SI{0.34}{\milli\electronvolt}; restricted to states within
\SI{1}{\electronvolt} of the alignment energy, which is the range that
decides a gap, a Fermi surface or a transport quantity, the median is at
most \SI{0.24}{\milli\electronvolt}. The largest single difference
anywhere in the window is \SI{0.64}{\milli\electronvolt}, in the
$d$-electron metal.

\begin{table*}[!t]
  \centering
  \caption{Band structure and density of states against the reference
           implementation, with both codes on one geometry per material. The
           gap columns report the fundamental gap, with ``overlap'' marking a
           metal. The median and RMS columns are the state-by-state
           band-energy difference over the band window described in the text,
           and the near-$E_{\rm F}$ column restricts that to states within
           \SI{1}{\electronvolt} of the alignment energy. The last column is
           the normalized density-of-states deviation over a
           \SI{6}{\electronvolt} window centred on the alignment energy.}
  \label{tab:bands}
  \footnotesize
  \setlength{\tabcolsep}{3.5pt}
  \begin{tabular}{llrrrrrr}
    \toprule
    Material & Character & \multicolumn{1}{c}{Gap (eV)} &
    \multicolumn{1}{c}{Gap (eV)} & Median & Near $E_{\rm F}$ & RMS & DOS \\
             &           & this code & reference & (meV) & (meV) & (meV) & dev. \\
    \midrule
    Si       & indirect-gap semiconductor & 0.6201 & 0.6201 & $-0.04$ & $-0.04$ & 0.05 & 0.00025 \\
    Na       & alkali metal               & overlap & overlap & $+0.06$ & $+0.04$ & 0.06 & 0.00002 \\
    Al       & simple metal               & overlap & overlap & $-0.02$ & $-0.03$ & 0.03 & 0.00012 \\
    Cu       & $d$-electron metal         & overlap & overlap & $+0.20$ & $-0.24$ & 0.34 & 0.00010 \\
    graphene & two-dimensional semimetal  & 0.0000 & 0.0000 & $+0.03$ & $+0.03$ & 0.04 & 0.00016 \\
    MoS\textsubscript{2} & two-dimensional semiconductor & 1.7691 & 1.7692 & $-0.10$ & $-0.03$ & 0.15 & 0.00138 \\
    \bottomrule
  \end{tabular}
\end{table*}

No material needs an exception to the table's single number, and the
window admits only states whose convergence the solver guarantees, so the
agreement is a property of the implementation and not of the sampling.

Running the same comparison with each code at its own relaxed geometry, which
is what an end-to-end comparison of two codes would measure, changes the
picture in one place. The four bulk materials relaxed to volumes agreeing
within \SI{0.07}{\percent} and their band differences barely move. Sodium
relaxed to volumes \SI{1.79}{\percent} apart, and its band difference rises
from \SI{0.1}{\milli\electronvolt} to \SI{13.2}{\milli\electronvolt} median.
The entire discrepancy in that material is the relaxation, and the band
solvers agree on it to a tenth of a milli\-electronvolt once they are given the
same cell. Neither number is an error: a soft alkali metal has a shallow
energy-volume curve, so a small difference in how the two codes converge the
geometry moves the volume, and the band structure moves with it.

Two further properties of these numbers need to be stated with them.

First, the absolute Fermi level of a gapped system is not a physical
observable, and none of the numbers above compare it. In silicon the two
implementations agree on the valence and conduction edges to
\SI{0.5}{\milli\electronvolt} and \SI{1.5}{\milli\electronvolt} and disagree by
\SI{91}{\milli\electronvolt} in the Fermi level itself, both values landing
inside the same \SI{0.62}{\electronvolt} gap.

Second, the geometry optimization of the two metals does not reach the force
criterion in the original runs: the alkali metal arrives at its step limit with
the forces still noisy, and the $d$-electron metal stops when the optimizer
trust radius falls to its floor. The common geometry removes that from the band
comparison, since the bands are then computed on a structure neither optimizer
produced, but it remains a property of the relaxation and is reported instead
of being smoothed over.

The sequence the six materials were carried through is the one a user runs.
A relaxation moves the ions and, when asked, the cell; a self-consistent step
converges the density at fixed geometry; a band-structure step reuses the
converged state and returns the eigenvalues along a path whose high-symmetry
points and coordinates are written with them; a density-of-states step reuses
that same state on the sampling the input names and returns the spectrum on a
uniform energy grid with the Fermi level recomputed from it. All
four are reachable from a control file, none of them requires a second code to
post-process the result, and each writes its own output and
not a single opaque state. The six materials exercise the branches that matter for a plane
wave code of this kind: a gap and a metal, states that disperse steeply and
states that do not, a two-dimensional cell with vacuum, and a layered cell
whose valence and conduction bands are separated by more than an
electronvolt.

\section{Properties of the software}
\label{sec:related}

Three properties define this code, and each carries a boundary worth stating
with it. They are summarised here and derived in the sections named, so that
a reader can see what the combination amounts to before working through the
detail.

The first is the implementation language. The code is written in Rust. The
argument for that choice rests on a guarantee rather than on speed: data
races are excluded at compile time, which is what a kernel that is threaded
over plane waves and bands would otherwise have to hold by convention. The
language is used the way the systems literature describes it, and the
constraints the project actually relies on are stated in
Section~\ref{sec:arch:rationale}. The boundary is that a memory-safe
language removes a class of defect and says nothing about whether the
physics or the arithmetic is right, which is what the validation apparatus
below is for.

The second is that the code is built around component-level validation,
with end-to-end regression as well. Section~\ref{sec:validation}
describes the corpus: numbered fixtures, each with the inputs, the reference
values and the script that regenerates them, covering individual kernels as
well as final energies. The argument for the form is that a numerical
change can then be attributed to a component instead of to a total. The
boundary is that a fixture proves agreement on the case it holds and not on
the cases it does not, and the corpus is a finite set of systems chosen by
the authors.

The third is that the physical surface is one combination and not a menu. A plane-wave basis with norm-conserving or
projector augmented-wave pseudopotentials, in the local-density or
Perdew--Burke--Ernzerhof approximation, periodic or molecular, serial or
distributed over three MPI axes with optional threading. The boundary is
that the combination is the whole of it: there is no spin-polarized,
non-collinear or spin--orbit path, no ultrasoft production path, no
accelerator execution, and no hybrid functional, and
Section~\ref{sec:overview:limits} states these together with the capabilities
they limit.

What the three add up to is a code whose scope is narrow enough to be
checked and whose checking is published with it. The claim is not that the
scope is wide; it is that within the scope the evidence can be re-run by
anyone who obtains the sources.

\section{Conclusions and outlook}
\label{sec:conclusions}

This paper describes \ares{}, a plane-wave density-functional-theory code
for periodic solids, written in Rust as a workspace of 22 crates. The code
covers norm-conserving and projector augmented-wave pseudopotentials, the
local-density and Perdew--Burke--Ernzerhof functionals, $\Gamma$-only and
explicit $k$-point sampling, smearing and mixing, forces, stress, fixed-cell
and variable-cell relaxation, band structures, densities of states and
dispersion corrections, and it runs serially or distributed over $k$-point
pools, band groups and plane-wave rows. Its numerical components are pinned
against an established plane-wave implementation one at a time, through a
corpus of numbered fixtures whose reference data is regenerated by a script
instead of being captured once.

The evidence for the code is of two kinds, and both are quantitative. Against
the reference implementation, driven from the same physical and numerical
inputs, the decomposed energy terms, densities, eigenvalues, forces, stress,
band structures and densities of states agree within stated thresholds, and
several of the comparisons reach the level at which two implementations can
be compared at all. Against all-electron reference data, the pseudopotentials
the code reads give a median delta factor of
\SI{0.696}{\milli\electronvolt\per atom} over all 65 paired
elements, with no element set aside. Where a
comparison does rest on a selection, the rule is stated with the table it
belongs to, because a benchmark whose dropped cases are invisible cannot be
checked.

Several limitations are current properties of the code, and a reader should
weigh them before choosing it.
\label{sec:overview:limits}
Two functionals are implemented, with no
spin-polarized, non-collinear or spin--orbit path, and the corresponding
input fields are parsed and rejected, not ignored. Occupations for the
self-consistent cycle are fixed or Gaussian; the other smearing schemes in
the grammar are not wired to the solver. The ultrasoft type is parsed and no
production path uses it. Stress carries a scope
narrower than that of the total energy, listed with the combinations it
covers. The absolute-charge consistency of
the augmented-wave augmentation is unresolved, and the density is
deliberately not renormalized to conceal it. Force and eigenvalue evidence
for large systems has not been established. Molecular dynamics is not
implemented and the DFT+U input block is reserved, so a control file that
names either is rejected. Accelerator execution is not
implemented, and continuous integration is not configured in the repository,
so the verification described here rests on recorded outputs of local runs.

Two things follow from where the code now stands. The first is that the validation apparatus is what makes the numbers in this paper
checkable, and no single result is: the fixtures and the scripts that regenerate
them are part of the artefact, so a reader who obtains the sources can
re-derive a comparison instead of taking it on trust. The second is portability. The numerical layer, the
parallel layer and the memory layer are separate crates whose dependencies
point one way, which keeps the solver, the process grid and the storage
backend independent of one another. That separation is what any later port
would build on. No port has been attempted here.


\section*{Code and data availability}
The \ares{} sources, the numbered cross-validation fixtures, the reference
data and the acceptance evidence described in Section~\ref{sec:validation}
are available from the corresponding author on request. The project name
carries the year; this paper cites the code as \ares{}, which is what the
acronym spells.



\bibliographystyle{unsrtnat}
\setlength{\bibsep}{1ex}
\bibliography{
  bib/references_all,
  bib/lit29_hylanemos,
  bib/lit30_pypwdft_memory,
  bib/lit31_inq_rttddft,
  bib/lit33_abinit2025,
  bib/lit34_fhi_aims_roadmap,
  bib/lit35_siesta_hybrids,
  bib/lit36_rust_julia_cise,
  bib/lit37_xrd_rust,
  bib/lit38_faer,
  bib/lit39_lamellar,
  bib/lit40_rustbelt,
  bib/lit51_pw_hybrid_sugon,
  bib/lit53_kssolv_toolbox,
  bib/r10_ares_anchor,
  bib/r11_ares_usage,
  bib/Blochl_1994,
  bib/Perdew_1996,
  bib/Perdew_1992,
  bib/Slater_1951,
  bib/Monkhorst_1976,
  bib/Ewald_1921,
  bib/Teter_1989,
  bib/Davidson_1975,
  bib/Broyden_1965,
  bib/Grimme_2006,
  bib/Grimme_2010,
  bib/Becke_Johnson_2005,
  bib/Lowdin_1950,
  bib/Lehtola_2018,
  bib/Lejaeghere_2016,
  bib/Blaha_2020,
  bib/Jollet_2014,
  bib/Hohenberg_1964,
  bib/Kohn_1965,
  bib/Payne_1992,
  bib/Mermin_1965,
  bib/Methfessel_1989,
  bib/Fletcher_1970,
  bib/Tiesinga_2021,
  bib/Stathopoulos_2002,
  bib/Fu_1983,
}

\end{document}